\documentclass[useAMS]{mn2e}
\usepackage{times}
\usepackage{amssymb}
\usepackage{rotating}
\input{epsf}

\title[QPOs and Lense-Thirring]
{Low frequency QPO spectra and Lense-Thirring precession}

\author[A. Ingram, C. Done, P. C. Fragile]
{Adam
Ingram$^{1}\thanks{E-mail:a.r.ingram@durham.ac.uk}$,
Chris Done$^{1}$ and P. Chris Fragile$^{2}$\\
$^1$Department of Physics, University of Durham, South Road,
Durham DH1 3LE, UK\\
$^2$Department of Physics and Astronomy, College of Charleston,
Charleston, SC 29424 USA\\
}

\date{Submitted to MNRAS}
\pagerange{\pageref{firstpage}--\pageref{lastpage}} \pubyear{2009}

\begin{document}

\topmargin = -0.5cm

\maketitle

\label{firstpage}

\begin{abstract}
We show that the low frequency QPO seen in the power density spectra
of black hole binaries (and neutron stars) can be explained by
Lense-Thirring precession. This has been proposed many times in the
past, and simple, single radius models can qualitatively match the
observed increase in QPO frequency by decreasing a characteristic
radius, as predicted by the truncated disc models. However, this also
predicts that the frequency is strongly dependent on spin, and gives a
maximum frequency at the last stable orbit which is generally much
higher than the remarkably constant maximum frequency at $\sim$~10~Hz
observed in all black hole binaries.  The key aspect of our model
which makes it match these observations is that the precession is of a
radially extended region of the hot inner flow. The outer radius is
set by the truncation radius of the disc as above, but the inner
radius lies well outside of the last stable orbit at the point where
numerical simulations show that the density drops off sharply for a
misaligned flow. Physically motivated analytic estimates for this inner
radius show that it increases with $a_*$, decreasing the expected
frequency in a way which almost completely cancels the expected
increase with spin, and ties the maximum predicted frequency to
around 10~Hz for all $a_*$. This is the first QPO
model which explains both frequencies and spectrum in the context of a
well established geometry for the accretion flow.

\end{abstract}

\begin{keywords}
X-rays: binaries -- accretion, accretion discs, black hole physics

\end{keywords}

\section{Introduction} \label{sec:introduction}

Stellar mass black hole binaries (BHB) are generically very variable
on timescales of 100-0.1s (0.01-10 Hz). A power spectral analysis of
the light curve shows that there are often quasi-periodic oscillations
(QPOs) superimposed on a broad band noise continuum. There are
several distinct types of QPO, but the most prominent is at low
frequencies. This QPO frequency {\em moves} from $\sim 0.1-10$~Hz in a
way which is tightly correlated with changes in the energy
spectrum of the source, with
both being dependent on the (average) mass accretion rate. At low
luminosities, the spectrum is dominated by a hard (photon index
$\Gamma < 2$) power law tail, peaking at ~100 keV (low/hard state) and
the QPO is a weak and broad feature at low frequencies. As the source
brightens, there is an increasing contribution from the disc at low
energies and the tail softens towards $\Gamma\sim 2$ (intermediate state)
as the QPO increases in frequency and becomes more coherent.
The tail can then either remain strong as the source flux
increases, giving a very high state with $\Gamma\sim 2.5$, and strong
QPO at $\sim 6-10$~Hz, or it can carry only a very small fraction of
the luminosity, giving the high/soft state with $\Gamma\sim 2.2$
and a weak or often non existent QPO
(e.g. the reviews by Remillard \& McClintock 2006; Done, Gierlinski \&
Kubota 2007, hereafter DGK07).

Despite this QPO being known for many decades (e.g. van der Klis
1989), its origin is still not understood. The frequency is far too
low to be associated with the orbital (Keplarian) timescale at the
last stable orbit, and the change in frequency also rules out any
association with a fixed radius. Instead, the correlation of the QPO
frequency with the spectral properties suggests another
possibility. The current best models for the hard spectra seen in the
low/hard state require that the cool, geometrically thin, optically thick
disc is truncated at some radius, $r_{o}$, which is larger than the
last stable orbit. The inner accretion flow instead forms a hot,
geometrically thick, optically thin configuration. The spectral
evolution within the low/hard state can be explained if the truncation
radius of the thin disc moves inwards with increasing mass accretion
rate, leading to greater illumination of the hot flow by the cool disc
photons, and hence to steeper spectra. This gives a physical basis for
the hard-soft transition when the disc finally reaches the last stable
orbit (Esin et al 1997; DGK07) and is supported by direct
observations of the disc receding in the soft-hard transition
(Gierlinski, Done \& Page 2008).  The inner radius of the thin disc
{\em moves} in these models, giving a possible origin for the moving
QPO frequency.

However, this only gives a qualitative association for the
characteristic frequency. A quantitative model linking the changing
truncation radius $r_o$ to the QPO frequency is required. Fundamentally,
there are 3 frequencies
associated with a given radius, $r$, namely that of a circular orbit
in the equatorial plane, $\nu_\phi$, a circular orbit inclined at
angle $\theta$ to the equatorial plane, $\nu_\theta$, and an
elliptical orbit in the equatorial plane with mean radius $r$,
$\nu_r$. In general relativity (as opposed to Newtonian gravity)
$\nu_r\ne \nu_\phi$, giving precession of the periastron of a radially
perturbed (elliptical) orbit.  This returns to its original azimuth on
the pericenter precession frequency $\kappa=\nu_\phi-\nu_r$. Similarly, tilted
orbits have $\nu_\theta\ne\nu_\phi$ if the black hole is spinning,
leading to a Lense-Thirring precession frequency of
$\nu_{LT}=|\nu_\phi-\nu_\theta|$.  Relativistic precession models
use these frequencies directly in order to produce the observed
QPOs, identifying the low frequency QPO (LF QPO) with Lense-Thirring
precession (Stella \& Vietri 1998; Stella et al 1999; Psaltis
\& Norman 2000, Fragile et al 2001). 

However, there are several problems with such an
identification. Firstly, this associates the frequency with the inner
edge of the thin (cool) disc, yet the QPO (and all the rapid
variability) are associated with the tail, not the disc (Churazov et
al 2001; Sobelewska \& Zycki 2006). Secondly, and more fundamentally,
it requires that all BHB are spinning. Given their birth in the
collapsing core of a massive star this is not surprising, with
estimates of $a_* <0.8-0.9$ (Gammie, Shapiro \& McKinney
2004). However, this allows for a {\em range} of spins in the BHB, as
is also suggested by observations (Davis, Done \& Blaes 2006; Shafee
et al 2006). The Lense-Thirring precession timescale depends strongly
on spin, so predicts that the same truncation radius in BHB of
different spin should give different QPO frequencies.  Yet
observations seem to show little difference in QPO frequencies from
object to object (Sobczak et al 2000; Pottschmidt et al 2003, Kalemci
et al 2004, Belloni et al 2005).  This makes any Lense-Thirring model
appear uncomfortably fine-tuned, especially in the light of tight
correlation between the low frequency break in the power spectrum and
QPO frequency which extends across both black hole {\em and} neutron
stars (Wijnands \& van der Klis 1999). The low frequency break is most
plausibly from the viscous timescale at $r_o$ (Psaltis \& Norman 2000;
DGK07), so does not depend on spin, unlike a Lense-Thirring model for
the QPO. Any range in spin between different objects (and neutron star
spins are {\em known} to range between $a_*=0.2-0.4$, while the black
hole spins can plausibly be significantly larger) should then give
rise to a large dispersion in the break-QPO relation, yet the data
limit this to less than a factor of 2 (Klein-Wolt \& van der Klis
2008).

Here we show how Lense-Thirring models can match the observations by
considering the physical interpretation of recent numerical
simulations. Specifically, we suggest that the \textit{shape} of the
warped, geometrically thick accretion flow which fills the region
inside $r_o$ affects the frequency at which it precesses. We discuss
how this precession can modulate the hard X-ray emission in order to
produce the observed energy dependence as well as frequency behaviour
of the QPO.

\section{Predictions for low frequency QPO spectra from Lense-Thirring
  precession} \label{sec:results}

\subsection{Point particle}
\label{sec:testmass}

The Lense-Thirring precession frequency
for pro-grade orbits in the limit of a small perturbation with respect
to the equatorial plane is 
\begin{eqnarray}
\nu_{LT}&=& |\nu _\phi - \nu _{\theta}| = \nu_\phi
\left[1-\sqrt{1-\frac{4a_*}{r^{3/2}}+\frac{3a_*^2}{r^2}}\right] 
\end{eqnarray}
where $\nu_\phi=c/[2\pi R_g(r^{3/2}+a_*)]$, $a_*$ is the dimensionless
spin per unit mass and $r$ is the dimensionless orbital
radius (in units of $R_g=GM/c^2$).  

\begin{figure} 
\centering
\leavevmode  \epsfxsize=8.5cm \epsfbox{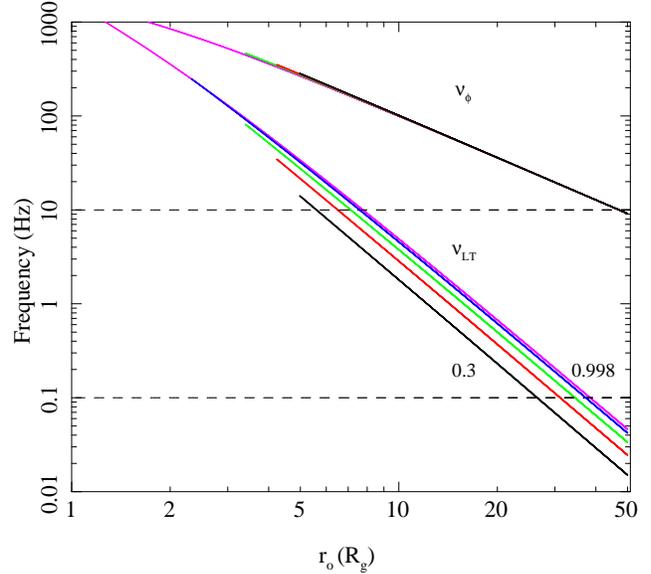}
\caption{Lense-Thirring precession frequency for a point particle and
Keplarian orbital frequency plotted against orbital radius. 
The solid black, red, green, blue and magenta lines depict $a_*=0.3$, $a_*=0.5$,
$a_*=0.7$, $a_*=0.9$ and $a=_*0.998$ respectively. The dashed lines represent the
limits of the observed range. Although Lense-Thirring precession gives
predictions closer to observation than Keplarian frequencies, the peak
frequency and $a_*$ dependence do not match observation.} \label{fig:point}
\end{figure}

We assume a black hole mass of $10M_{\sun}$ throughout this paper.
Figure \ref{fig:point} shows the precession frequency for a variety of
black hole spins from $0.3<a_* < 0.998$ as a function of radius down
to the last stable orbit, $r_{lso}$. This clearly shows that the highest
frequency predicted is heavily dependent on spin, and that these are
well in excess of the $\sim 10$Hz maximum observed QPO frequency for
$a_*>0.3$. The corresponding Keplarian frequencies (upper lines) plotted
for comparison trace out even higher values. 

\subsection{Solid disc with inner radius at the last stable orbit}
\label{sec:torlso}

\begin{figure}
\centering
\begin{turn}{-90}
\leavevmode  \epsfxsize=3.0cm \epsfbox{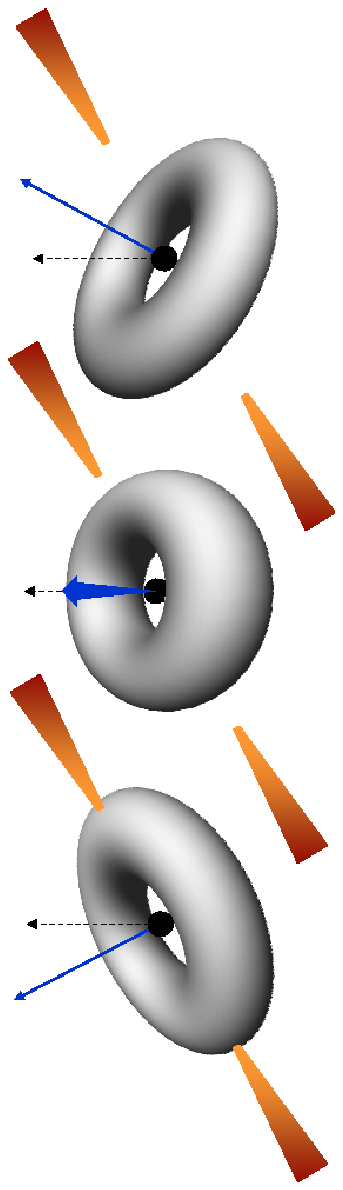}
\end{turn}
\caption{Schematic diagram of the geometry considered. The inner flow
(grey with blue angular momentum vector) precesses about the black hole
angular momentum vector whilst the outer disc (red/orange) remains aligned
with the binary partner. The flow extends between $r_i$ and $r_o$.}
\label{fig:schem}
\end{figure}

The simple estimates in the previous section assumed single particle
orbits at the truncation radius of the thin disc, $r_o$. However, the
energy dependence of the QPO clearly associates it with the hot flow
rather than the disc. Thus we consider Lense-Thirring precession of
the geometrically thick, hot flow interior to the truncated disc as
illustrated in Figure \ref{fig:schem}. Fragile et al. (2007) estimate
the associated frequency assuming that the black hole
torque from the misalignment makes the entire flow precess as a
solid body between an inner and outer radius, $r_i$ and $r_o$ (again
scaled in units of $R_g$ to make them dimensionless). This gives
\begin{equation}
\nu_{prec} = \frac{(5-2\zeta)}{\pi(1+2\zeta)}
\frac{a_*[1-(r_i/r_o)^{1/2+\zeta}]}
{r_o^{5/2-\zeta}r_i^{1/2+\zeta}
[1-(r_i/r_o)^{5/2-\zeta}]}\frac{c}{R_g}
\label{eqn:tprec}
\end{equation}
where the moment of inertia of the disc is calculated assuming a
surface density of the form $\Sigma=\Sigma_o (r/r_i)^{-\zeta}$.  Classical
advection dominated accretion flows give $\zeta=0.5$, while thin discs
have $\zeta \sim -0.5$, and the numerical simulations give $\zeta\sim
0$. We choose  $\zeta=0$, but note this makes less than a factor of 2
difference from the other prescription for the resultant QPO frequency 
even at the largest radii, and that this difference decreases
monotonically as $r_o$ decreases. 

\begin{figure}
\centering
\leavevmode  \epsfxsize=8.5cm \epsfbox{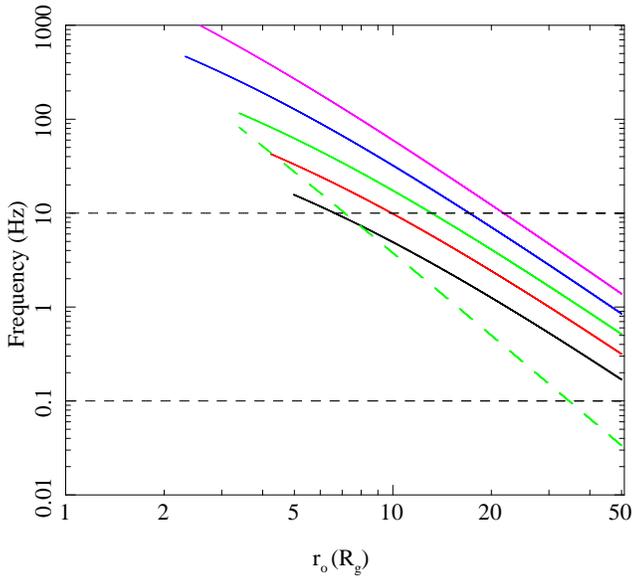}
\caption{Precession frequency of an inner flow of varying outer radius.
The solid black, red, green, blue and magenta lines represent spin values
of $a_*=0.3$, $a_*=0.5$, $a_*=0.7$, $a_*=0.9$ and $a_*=0.998$ respectively.
The green dashed line represents a point particle for $a_*=0.7$. The minimum
radius is the last stable orbit as a function of spin. We see that, as in
the case of point particle Lense-Thirring, the peak frequency is both higher
than observed values and has too strong a spin dependence.} \label{fig:torlso}
\end{figure}

Figure \ref{fig:torlso} shows the precession frequency plotted against
$r_o$ for a number of spins with $r_i=r_{lso}$. These frequencies are
always higher at a given $r_o$ as the effective radius is a surface
density weighted average from $r_i$ to $r_o$
We still, however, see the
same two problems encountered in section \ref{sec:testmass}, namely,
that the peak frequency is too high and varies too strongly with spin.

\subsection{Inner radius}
\label{sec:beta}

So far we have considered a flow with its inner radius at the last
stable orbit. Instead, the precession timescale is set by where the
surface density drops significantly, as the region interior to this
will not contribute significantly to the moment of inertia.  Full
general relativistic simulations of the magneto-rotational instability
(MRI, the underlying source of the stresses which transport angular
momentum) show that this drops sharply at around $1.5\times r_{lso}$
(e.g. Fig 4. in Krolik, Hawley \& Hirose 2005) for thick flows aligned
with the black hole spin.

\begin{figure}
\centering
\leavevmode  \epsfxsize=8.5cm \epsfbox{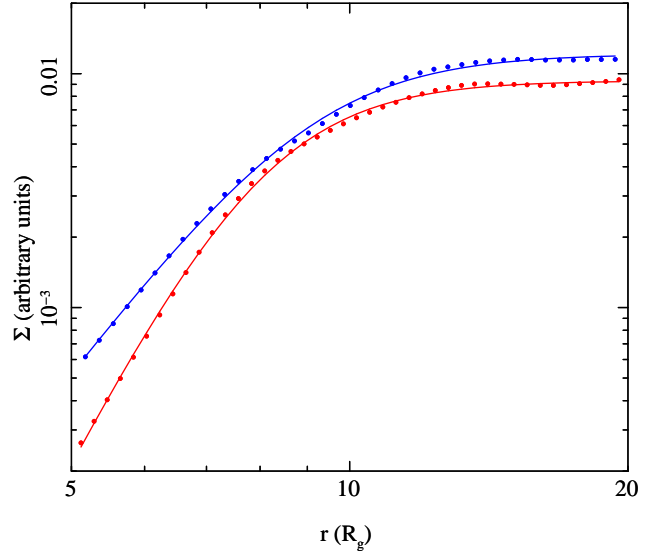}
\caption{Surface density as a function of radius recovered from
numerical simulations of a misaligned flow (Fragile et al 2007) with $a_*=0.5$
(red) and $a_*=0.9$ (blue). Data points have been fit by a double law
which breaks at $r_i$. We find $r_i(a_*=0.5)\sim8$ and $r_i(a_*=0.9)\sim9$.}
 \label{fig:sigma}
\end{figure}

However, we are considering Lense-Thirring precession so the key issue
is that the flow is {\em misaligned}. The extra torques on the flow
give extra contributions to the stresses. Simulations (e.g. Fragile et al 2007)
have shown this to increase the inward velocity, and therefore decrease
the density of the flow. Figure  \ref{fig:sigma} shows the surface density profile
obtained from two simulations, both of a flow misaligned by $15^o$ but with
differing black hole spin. The blue points are for $a_*=0.9$ (Fragile et al 2007)
and the red points are for $a_*=0.5$ (Fragile et al 2009). We have fit the
data with a smoothly broken power law function 
$\Sigma_o x^\alpha/(1+x^\gamma)^{(\zeta+\alpha)/\gamma}$ where
$x=r/r_i$.  This gives $x^\alpha$ and $x^{-\zeta}$ for $r<<r_i$ and
$r>>r_i$, respectively, while $\gamma$ controls the sharpness of the
break.  We fix $\zeta=0$ (see Section \ref{sec:torlso})
and obtain $r_i\sim 9$ for $a_*=0.9$ and $r_i\sim8$ for $a_*=0.5$, both of
which are significantly larger than $r_{lso}-1.5~r_{lso}$ for untilted
flows.

Ideally, we would now like to re-plot figure \ref{fig:torlso} using
the inner radius for a misaligned flow. However, we only have two
simulation points for $r_i$ which is clearly inadequate for our
purposes. We therefore make an analytical approximation in the next section
in order to address this point.

\subsubsection{Solid disc with inner radius set by bending waves}
\label{sec:incoherent}

The additional torques will be strongest where the flow is most
misaligned, so these should track the {\em shape} of the flow.  This
is set by bending waves, which communicate the warp and twist in
initially circular and coplanar orbits, against viscous
damping. Analytic approximations to the resulting shape can be
calculated assuming linear perturbations in an initially thin disc
(e.g. Ferreira \& Ogilvie 2008).  The global structure then depends on
the ratio of the viscosity parameter, $\alpha$, relative to the disc
semi-thickness, $H=hR_g$. For $\alpha > h/r$, warped disturbances via
Lense-Thirring precession are propagated by viscous decay which eventually
drags the inner disc into alignment with the black hole spin, while the
outer disc aligns with the orbital plane of the companion star
(Bardeen \& Peterson 1975, King et al 2005).  The Bardeen-Petterson
transition radius can be roughly defined as the point where viscosity
can no longer propagate warps in the disc outward quickly enough to prevent a
twisting of the disc due to differential precession.

However, we are considering instead a hot inner flow which is
geometrically thick so $\alpha < h/r$.  In this case, the warp is
propagated via bending waves. The local sound crossing timescale is shorter
than the precession timescale throughout the flow allowing the material to
be strongly coupled by pressure waves. Consequently, the flow
precesses as a solid body (Fragile et al 2007) with its shape influenced by the
bending waves (Ferraria \& Ogilvie 2008; Pringle 1992, Lubow et
al. 2002). Undamped bending waves have a characteristic wavelength of
\begin{equation}
\lambda\sim {\pi r^{9/4}\over (6a_*)^{1/2} } \Bigl( {h\over r}
  \Bigr).
\label{eqn:lambda}
\end{equation}
These waves are, therefore, smooth at large radii and oscillatory at
small radii due to the strong $r$ dependence of the wavelength. Figure
12 in Fragile et al (2007) and Figure 10 in Fragile et al (2009)
show the tilt angle of the flow at varying
radii for $a_*=0.9$ and $0.5$, respectively. 
This tilt angle increase dramatically at small radii
in a manner similar to that of the bending waves. It could be that this
rapid change in disc tilt gives rise to additional stresses which lead
to the observed drop off in surface density. It is encouraging that
Figure 13 in Fragile et al (2007) seems to support this assertion as
it shows that the viscosity parameter of the disc, $\alpha$, increases
rapidly at small radii.

The largest radius at which the rate of change of disc tilt is
significant is $r \sim \lambda /4$ i.e. at $r_i\sim 2.5 (h/r)^{-4/5}
a_*^{2/5}$ (using equation \ref{eqn:lambda}) as this is the first
point at which the bending waves have room to turn over.  A more
rigorous treatment by Lubow et al (2002) gives $r_i\sim 3.0
(h/r)^{-4/5} a_*^{2/5}$.  Both of these expressions give $\sim 10$ and
$8$ for $a_*=0.9$ and $0.5$, respectively, for $h/r=0.2$, in agreement
with the simulations (see Section \ref{sec:beta}).

\begin{figure}
\centering
\leavevmode  \epsfxsize=8.5cm \epsfbox{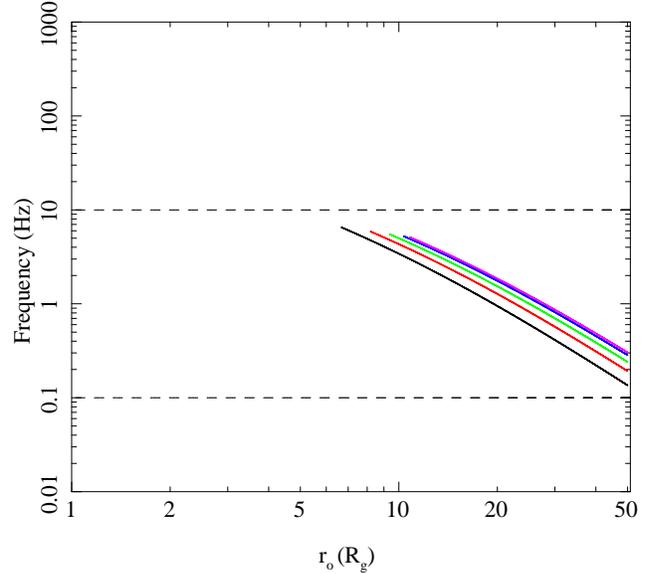}
\caption{Precession frequency versus outer radius of a
hot flow with scale height $h/r=0.2$ and inner radius set by the
bending wave region $r_i=3.0(h/r)^{-4/5} a_*^{2/5}$ (Lubow et al 2002)
for spins of $a_*=0.3$ (black), $a_*=0.5$ (red), $a_*=0.7$ (green),
$a_*=0.9$ (blue) and $a_*=0.998$ (magenta). The expected increase in QPO
frequency with $a_*$ is mostly cancelled out by the increased radial
extent of the bending wave region, and the maximum QPO frequency is
$\sim10Hz$, as observed.} 
\label{fig:nuprecro}
\end{figure}

Figure \ref{fig:nuprecro} shows the precession frequency recalculated
assuming the inner radius as above.  We see that the expected decrease
in QPO frequency with spin is offset by the increase in inner radius
with spin. Counter-intuitively, the QPO probes smaller radii in the
flow for lower black hole spins! Figure \ref{fig:nuprecro} is in fact
remarkably like the observed data in that it predicts a maximum
frequency of $\sim 10$~Hz for all spins considered here ($a_*>0.3$).
It also predicts the frequency to be mostly dependent on the outer
radius of the flow, not spin, which allows the QPO frequency to
tightly correlate with any other frequency picked out by this radius,
e.g. the low frequency break in the broad band power spectrum (Psaltis
et al 1999; Wijnands \& van der Klis 1999; Psaltis \& Norman 2000).

This is a very encouraging result, but we caution that many more
simulations are needed to quantify the behaviour the inner radius as a
function of spin, and to assess the effect of misalignment angle (both
current simulations are for $15^\circ$). Such simulations also mean
that the simplified form of the surface density profile in equation 2
can be replaced by the {\em observed} precession frequency of the
flow.  However, the two current simulations show the drawback of this
approach as this is also sensitive to the {\em outer} boundary
condition. Our model sets the outer radius of the precessing flow by
the inner edge of the truncated disc. The flow can only freely
precess in the region where there is no thin disc
blocking the mid-plane. Instead, the current simulations only include
the hot flow, and its effective outer radius is larger for the
$a_*=0.5$ run than in the $a_*=0.9$ and the precession frequency
directly observed from the simulations is roughly a factor of two
higher for $a_*=0.9$ than for $0.5$. Thus the simulations need also to
include an outer boundary condition in order to properly explore
parameter space, and to consider the additional torque on the flow
from the interaction between the thin disc and hot flow which adds
a great deal of complexity.

\section{Discussion} \label{sec:discussion}

The Lense-Thirring frequency of the inner flow precessing as a solid
disc does not match observed LF QPO frequencies if we assume the inner
radius of the flow is set by the last stable orbit.  However, recent
numerical simulations show that the surface density profile of a
misaligned flow drops substantially at radii which are significantly
larger than $r_{lso}$ for $a_*=0.9$ and $a_*=0.5$. We postulate that
this radius is set by the shape of the bending waves which distort the
disc. This radius increases with $a_*$ in a way that counteracts most of the
expected increase in QPO frequency with spin at a given $r_o$.
This results in a maximum value of $\sim 5-10Hz$ for a $10M_\odot$
black hole of almost any spin, as observed.

Clearly this conclusion depends on the outcome of future numerical
simulations. It also depends on the flow being misaligned!
The Bardeen-Peterson effect dictates that a misaligned thin disc will
gradually align with the black hole at small enough radii (Bardeen \&
Peterson 1974). Most analytical estimates predict that the disc should
be more or less aligned at typical values of the truncation radius
(e.g. Fragile et al 2001). This, therefore, implies that the flow should
be aligned if most of the material for the hot flow accretes through the
outer disc which, in turn, implies that it shouldn't precess! However,
the thin disc alignment should be rather different for a truncated thin
disc. We intend to explore this effect in future work.

These caveats aside, we have a very attractive model for the
origin of the low frequency QPO in black hole binaries. This is made
even more compelling as it ties the QPO to the hot flow, so should
directly modulate the Comptonised tail, as observed, even though the
outer radius $r_o$ is determined by the thin disc.


There are several processes which can imprint the modulation on the
spectrum. The flow is translucent (optical depth, $\tau\sim 1$) so there
can be weak projected area effects. More importantly, the flow can
self-occult causing a dip in the flux of maximum depth $exp(-\tau)$
when the flow is aligned with the observer's line of sight.
There should also be a variation in the number of seed photons from the
disc irradiating the flow which, for example, will give maximum flux when
the flow is maximally misaligned with the disc. Relativistic effects can
also contribute to the modulation (Schnittman, Homan \& Miller 2006;
Schnittman \& Rezzolla 2006)

These effects will give a stronger modulation for higher inclination
angles, and higher optical depth. There is observational evidence for
both of these, with a compilation of BHB showing that the maximum QPO
r.m.s. strength increases with inclination, and with mass accretion
rate i.e. optical depth of the hot flow (Schnittman, Homan, \& Miller
2006). This can also explain why the QPOs appear stronger on the
hard--to--soft transition during the rapid rise to outburst than on
the soft--to--hard transition on the decline. The
hysteresis effect (plausibly caused by the rapid rise driving the disc
into a non-steady state configuration: DGK07) means that the
luminosity during the rise is higher, so the mass accretion rate
through the hot flow and hence its optical depth are both also larger,
giving stronger modulation.

This all fits well with the observations that broader, weaker QPOs are seen
in the low inclination systems such as 4U~1543-47 (Schnittman, Homan
\& Miller 2006). Our model also {\em predicts} that Cyg X-1 should have the
weakest QPOs, as observed, as it is both at low inclination {\em and}
is stable to the Hydrogen ionisation instability which drives the
enhanced optical depth seen during hysteresis (DGK07).

The physical processes in our model are scale invariant, so predict that the
frequencies for a given black hole spin, $a_*$, depend linearly on
mass, as generally assumed (Vaughan \& Uttley 2005; Gierlinski et al
2008; Middleton et al 2008). The BHB alone probably span
$6-14 M_\odot$ (Remillard \& McClintock 2006), so this predicts
a factor of 2.3 variation in frequency which may be detectable. 

\section{Conclusions}

Lense-Thirring precession of a radially extended section of the hot
inner flow in the truncated disc models can match the properties of
the low frequency QPO in BHB. The outer radius of the precessing flow
is set by the truncation radius of the cool disc. This sweeps inwards
as the source makes a transition from the low/hard to high/soft state
(DGK07). The surface density of a misaligned flow drops off at an
inner radius greater than the last stable orbit (Fragile et al
2007). The expected increase in QPO frequency with spin is mostly
counteracted by the increasing inner radius in our (albeit
speculative) models for $r_i$.  This gives a maximum predicted QPO
frequency of 6-10~Hz irrespective of spin, as observed in all BHB.
Thus while the QPO mechanism fundamentally depends on black hole spin,
the behaviour of the inner radius of the hot flow means that it does
{\em not} give a simple diagnostic of $a_*$.

The QPO arises from the hot flow, so naturally modulates the hard
X-ray flux through a combination of self occultation, projected area
and relativistic effects. These become stronger as a function of 
inclination and optical depth, as observed. 

This gives the first mechanism for the QPO which predicts both its
frequency and spectral behaviour, and embeds it firmly in the models
for the accretion flow and associated jet. If confirmed by further
numerical simulations, this solves the 20 year mystery of
how these characteristic frequencies arise in the accretion flow. 

\section*{Acknowledgements}

CD and PCF thank Axel Brandenburg and Nordita (Stockholm) for support to attend the
workshop on 'Turbulence and oscillations in accretion discs' which
provided much of the motivation for this work and many useful
conversations with Omer Blaes and Barbera Ferreira.
ARI acknowledges support from an STFC studentship. The authors would like to
thank the referee for useful comments which significantly improved this paper.


\label{lastpage}

\end{document}